\documentclass[12pt]{article}
\title{
\vspace{-3mm} \rightline{\small IFUP-TH 2003/15} \vspace{8mm} \bf
SUSY 3D Georgi - Glashow model at finite temperature}
\author{
Dmitri Antonov \thanks{
E-mail: {\tt antonov@df.unipi.it}}
\thanks{Permanent address:
ITEP, B. Cheremushkinskaya 25, RU-117 218 Moscow, Russia.}\\
{\it INFN-Sezione di Pisa, Universit\'a degli studi di Pisa,
Dipartimento di Fisica,}\\
{\it Via Buonarroti, 2 - Ed. B - I-56127 Pisa, Italy}\\
Alex Kovner\thanks{ E-mail: {\tt akovner@plymouth.ac.uk}} \\{\it
Department of Mathematics and Statistics, University of
Plymouth,}\\{\it Drake Circus, Plymouth PL4 8AA, UK}}

\date{}
\begin{document}


\maketitle \vspace{1mm} \centerline{\bf {Abstract}} \vspace{3mm}
\noindent We study the finite-temperature properties of the
supersymmetric version of (2+1)D Georgi-Glashow model. As
opposed to its nonsupersymmetric counterpart, the parity symmetry
in this theory at zero temperature is spontaneously broken by the
 bilinear photino condensate. We find that as the temperature is
raised, the deconfinement and the parity restoration occur in this
model at the same point $T_c=g^2/8\pi$.  The transition is
continuous, but is not of the Ising type as in nonsupersymmetric
Georgi-Glashow model, but rather of the
Berezinsky-Kosterlitz-Thouless type as in $Z_4$-invariant spin
model.

\vspace{5mm}

\newpage

During the last two years it has been realized that the
finite-temperature structure of weakly interacting 3D non-Abelian gauge
theories can be analysed exactly. It has also been found that these
theories exhibit many phenomena similar to what one expects to
find in 4D QCD and thus are useful solvable toy models for the
study of the deconfining dynamics at finite temperature. In
particular, various properties of the deconfining phase transition
in the $SU(2)$ Georgi-Glashow model have been understood. The
order of the phase transition as well as the universality class
have been established explicitly without recourse to universality
arguments, and the dynamics of the phase transition was given a
simple interpretation in terms of restoration of the magnetic symmetry~\cite{2}.
In subsequent work the effects of instantons at high
temperature have been understood in detail, the dynamics of the
deconfining transition has been related to the properties of
confining strings, and the analysis has also been extended to the
$SU(N)$ Georgi-Glashow model at $N>2$~\cite{gg2}.  Also some
interesting analogies between the mechanism of the deconfining
transition in 2+1 dimensions and the chiral-symmetry restoration
in QCD have been suggested~\cite{gg3}. These results have been
reviewed and summarized in \cite{kk}. Recently, it has also
been shown that the presence of heavy dynamical fundamental quarks
turns the second-order deconfining transition into analytic but
rather fast crossover~\cite{dkn}. Finally, the
effects of variability of the Higgs-field mass, as well as the effects of
light fundamental fermions on the monopole interaction have been studied in Ref.~\cite{antonov}.

In the present note, we continue this line of investigation and consider
the supersymmetric generalization of the 3D Georgi-Glashow model
at finite temperature. The interest in this model is that it
contains adjoint fermions whose masslessness is protected by the
discrete parity symmetry. At zero temperature, the parity is
spontaneously broken via a nonvanishing photino condensate. Thus,
at finite temperature one may anticipate two phase transitions --
one related to the vanishing of the photino condensate and the
other one due to deconfinement. These two transitions could either
be distinct and happen at different temperatures, or could
coincide. In this respect the model is similar to QCD with adjoint
quarks, where a similar question can be asked about the
(non-)coincidence of deconfinement and restoration of discrete
chiral symmetry.

The Lagrangian of supersymmetric Georgi-Glashow (SGG) model
contains the bosonic fields of the non-supersymmetric
Georgi-Glashow (GG) model, that are the light photon, the heavy
$W^\pm$ vector bosons and the massive Higgs field, as well as
their superpartners -- photino, winos and Higgsino. It was shown
in ~\cite{ahw} that just like in the GG model the monopole effects
render the photon massive, although the mass in this case is
parametrically smaller, since it is due to the contribution from a
two-monopole sector, rather than a single-monopole sector as in
the GG model. Since supersymmetry is not broken, the low-energy
sector of the theory contains in addition to the photon, the light
photino and is described by the supersymmetric sine-Gordon model.
Its Euclidean action in the superfield notation reads~\footnote{We
adopt here the notations of Ref.~\cite{MZ}, in particular $\int
d^2\theta\bar\theta\theta=1$.}

\begin{equation}
\label{0}
S=-\int d^3xd^2\theta\left[\frac12\Phi\bar D_\alpha D_\alpha\Phi+\xi
\cos\left(g_m\Phi\right)\right].
\end{equation}
In this equation, the scalar supermultiplet and supercovariant
derivatives have the form

\begin{equation}
\Phi({\bf x},\theta)=\chi+\bar\theta\lambda+\frac12\bar\theta\theta F,~~
D_\alpha=\frac{\partial}{\partial\bar\theta_\alpha}-(\hat\partial\theta)_\alpha,~~
\bar D_\alpha=\frac{\partial}{\partial\theta_\alpha}-(\bar\theta\hat\partial)_\alpha.
\end{equation}
Here, $\chi$ is the dual-photon field (real scalar), $\lambda$ is
the photino field, which is the two-component Majorana spinor
($\bar\lambda=\lambda^T\sigma_2$), $F$ is an auxiliary scalar
field, $\hat\partial\equiv\gamma_i\partial_i$, and the Euclidean
$\gamma$-matrices coincide with the Pauli matrices:
$\vec\gamma=\vec\sigma$. The "magnetic coupling" $g_m$ is related
to the gauge coupling of the SGG model as $g_{m}= 4\pi/g$ and has
dimensionality $[{\rm mass}]^{-1/2}$. The coefficient $\xi$ is the
monopole fugacity of dimensionality $[{\rm mass}]^{2}$ and is
exponentially small. The interaction term in Eq.~(\ref{0}) is
frequently understood as normal ordered. In this case, the
fugacity in terms of the mass of the $W$-bosons (in the BPS limit)
is $\xi\propto \exp\left(-4\pi M_W/g^2\right)$~\cite{1, ahw}. We
will find it however more convenient to use the non-normal ordered
form of the interaction. In this case $\xi$ is not as small, but
still has an exponential smallness $\xi\propto\exp(-S_{\rm
core})$, where $S_{\rm core}$ is the action of the monopole core.
The action $S_{\rm core}$ is the contribution to the monopole
action due to heavy particles -- $W$-bosons, Higgs and their
superpartners, and is a number of order $O\left({M_W\over
g^2}\right)$. All results of the present note are valid to the
lowest order in this parameter.

In the component notations, the action~(\ref{0}) can be readily
rewritten up to a constant as (cf. also Ref.~\cite{ahw})

\begin{equation}
S=\int
d^3x\left[\frac12(\partial_i\chi)^2-\frac{1}{2}\bar\lambda\hat\partial\lambda-
{g_m^2\zeta\over 2} \left(V^2+V^{*2}\right)\bar\lambda\lambda-
{(g_m\zeta)^2\over 2}\left(V^4+V^{*4}\right)\right],
\label{action}\end{equation}
where $\zeta=\xi/4$, and
we have defined the
vortex operator
\begin{equation}
V(x)=\exp\left(i{2\pi\over g}\chi\right).
\end{equation}
Just like the GG model, the model~(\ref{action}) has a magnetic $Z_2$ symmetry~\cite{2}.
It is easiest recognized by its action on the order
parameter, the vortex field:

\begin{equation}
V(x)\rightarrow -V(x).
\label{M}
\end{equation}

Besides the magnetic $Z_2$ symmetry, the effective action~(\ref{action})
has an additional discrete parity symmetry
inherited from the full SGG action,
\begin{equation}
V(x_1,x_2,x_3)\rightarrow iV(-x_1,x_2,x_3),\ \ \ \ \
\lambda(x_1,x_2,x_3)\rightarrow\sigma_3\lambda(-x_1,x_2,x_3).
\label{P}\end{equation} The photino mass term is odd under the
parity transformation~(\ref{P}). Thus, the photino can acquire a
mass only if the parity is spontaneously broken. It is indeed easy
to see that this is the case in the effective
Lagrangian~(\ref{action}). The potential of the dual-photon field
is minimized at $\left<\chi\right>=0$, or $\left<V\right>=1$. The
real expectation value of $V$ violates both the magnetic $Z_2$
symmetry and the parity symmetry. The spontaneous breaking of the
magnetic $Z_2$ symmetry is synonymous with
confinement~\cite{kovner}.  The breaking of parity results in the
non vanishing photino condensate
\begin{equation}
\left<\bar\lambda\lambda\right>\sim g_m^2\zeta \Lambda.
\end{equation}
The ultraviolet cutoff in the effective theory~(\ref{action}) is
of course of the order of $M_W$ -- the mass of the $W$ bosons.

The breaking of parity leads to the non vanishing photino mass
$m=2g_m^2\zeta$. On the classical level, the mass of the photon
can be read off from the photon self-interaction term. The photon
and the photino are of course degenerate as a consequence of an
unbroken supersymmetry.  It is in fact quite amusing to see how
this degeneracy is preserved on the quantum level. The simplest
loop corrections are those due to summation of "bubble diagrams",
or "normal-ordering" corrections. Taking those into account, the
photino mass becomes
\begin{equation}
m=g_m^2\zeta \left<V^2+V^{*2}\right>=2g_m^2\zeta {\rm e}^{-{2\pi\over g^2}\Lambda}.
\end{equation}
On the other hand, when we examine the normal-ordering corrections
to the photon self-interaction, the result is
\begin{equation}
{(g_m\zeta)^2\over 2}\left(V^4+V^{*4}\right)={(g_m\zeta)^2\over 2} {\rm e}^{-{8\pi
\over g^2}\Lambda}:\left(V^4+V^{*4}\right):~ .
\end{equation}
Thus, on the quantum level the self-interaction term gives the
contribution to the photon mass which is exponentially smaller
than the mass of the photino! This of course does not mean that
the supersymmetry is broken, but merely that the main contribution
to the photon mass comes from the diagrams containing the photino.

To calculate the photon mass to the order $O(\zeta)$, we have to
find the effective potential to the order $O(\zeta^2)$.
Integration over the photino yields the following
$O(\zeta^2)$-contribution to the effective action:
\begin{equation}
-\frac{(g_m^2\zeta)^2}{32\pi^2}\int d^3x\int
d^3y{1\over(x-y)^4}\left[V^2(x)+V^{*2}(x)\right]\left[V^2(y)+V^{*2}(y)\right].
\end{equation}
The effective potential is obtained by further integration over
the field $\chi$, keeping the zero-momentum mode of $\chi$ as a
fixed background. Since the integral is dominated by the distances
$|x-y|$ much smaller than the inverse photon mass, we can with the
exponential accuracy take $\chi(x)$ as a free massless field. We
then get
\begin{equation}
U_{\rm eff}=-\frac{(g_m^2\zeta)^2}{32\pi^2}{\rm e}^{-{4\pi\over
g^2}\Lambda}\left[\int d^3(x-y){1\over(x-y)^4}{\rm e}^{-{4\pi\over g^2
|x-y|}}\right]\left[V^4+V^{*4}\right]+{\rm const.}
\label{integral}\end{equation} The integral over $x-y$ is easily
performed with the expected result
\begin{equation}
U_{\rm eff}=-{(g_m\zeta )^2\over 2}{\rm e}^{-{4\pi\over
g^2}\Lambda}\left(V^4+V^{*4}\right),
\end{equation}
so that indeed the equality of the photon and photino masses is
preserved.

An interesting property of this calculation is that the main
contribution to the effective potential in the integral~(\ref{integral})
comes from the distances of order $O\left({1\over
g^2}\right)$. The contribution of large distances is suppressed by
the photino propagator, while the short distances are cut off by
the photon propagator in the exponential in Eq.~(\ref{integral}).
The saddle point of the integral in Eq.~(\ref{integral}) is in fact
at $|x-y|={2\pi\over g^2}$. The reason this is of some interest
is that as we know from the study of the GG model at
finite temperature, it is exactly in the range of temperatures
$T\sim g^2$, where the interesting phase transitions occur.

Let us now turn to the study of the model at finite temperature.
We will be mostly interested in temperatures of order $g^2$. Since
this is much higher than both the photon and photino masses, the
proper way to proceed is via dimensionally reduced theory of the
zero Matsubara mode. To derive it, we have to integrate out the
fermions and the nonzero Matsubara modes of the photon field.
Technically, to the order $O\left(\zeta^2\right)$ this calculation
is very similar to the one just performed. One keeps the zero
Matsubara mode of $\chi$ as fixed external background and
integrates over the rest of the degrees of freedom. Omitting the
exponentially suppressed terms, the result is
\begin{equation}
S=  \int d^2x\left[\frac\beta2(\partial_i\chi)^2- {\bar\zeta
^2\over 2}\left(V^4+V^{*4}\right)\right]\label{dimred}\end{equation} with
\begin{equation}
\bar\zeta^2= \zeta^2\beta g_m^2 {\rm e}^{-{4\pi\over
g^2}\Lambda}\int d^3x D_\beta^2(x){\rm e}^{-{16\pi^2\over g^2}
G_\beta(x)}.
\end{equation} Here, $\beta=1/T$, $D_\beta(x)$ is the finite-temperature
massless-fermion propagator, and $G_\beta(x)$ is the finite-temperature
massless-boson propagator with the contribution of
zero Matsubara frequency subtracted.

The exact value of $\bar\zeta$ is not important. It is however
clear that it is positive for any finite temperature and at
temperatures of interest it is in fact parametrically of the same
order as at zero temperature. The issue of sign is important for
the following reason. If $\bar\zeta^2$ were to change sign at some
temperature $T^*$, the classical vacuum of the potential in
Eq.~(\ref{dimred}) would shift from $\langle V\rangle=\pm 1$ to
$\langle V\rangle={1\pm i\over \sqrt 2}$. At this new value, we
would have $\langle V^2+V^{*2}\rangle=0$, and parity would be
restored. This would not be a deconfining transition, since
$\langle V\rangle\ne 0$, and thus the magnetic $Z_2$ symmetry would remain
broken. The change of sign of $\bar\zeta^2$ thus would mean that
the deconfining phase transition is preceded by the parity
restoration.

However, it is easy to see that this does not happen in our model.
The integral in $\bar\zeta^2$ is explicitly positive, since for any
temperatures $D_\beta(x)$ and $G_\beta(x)$ are both real functions. The
use of thermal propagators effectively limits the integration
region to $|x|<2\pi\beta$. Since, as we noted above, the main
contribution to the integral comes from the distances $|x|\sim
2\pi/g^2$, this means that for temperatures of interest ($T\sim
g^2$) the finite part of the relevant integration region
contributes, and thus the integral parametrically has the same
value as at zero temperature.

As discussed in \cite{2,kk}, to study the deconfining phase
transition  one cannot neglect the heavy charged degrees of
freedom. The thermal excitation of $W$ bosons (and their
superpartners) leads to appearance of extra operators in the high-temperature
effective action.  As discussed in detail in
\cite{2,kk}, the most important such operator is the adjoint
Abelian Polyakov line, which is the variable dual to the vortex
operator. The respective complete Lagrangian is
\begin{equation} S=  \int d^2x\left[\frac\beta2(\partial_i\chi)^2-
{\bar\zeta^2\over 2}\left(V^4+V^{*4}\right)- \mu\left(P^2+P^{*2}\right)
\right],\label{comact}\end{equation} where
\begin{equation} P=\exp\left(i{\tilde\chi\over 2}\right) \end{equation}
and $\tilde\chi$
is the field dual to $\chi$:
 \begin{equation} i\partial_{\mu}\tilde\chi= {g\over  T}
\epsilon_{\mu\nu}\partial^\nu \chi. \label{chid} \end{equation}
The last relation is valid modulo quantized discontinuities in the
phase $\chi$ and $\tilde\chi$, and  more properly
\begin{equation}
iP^*\partial_\mu P ={g^2\over 4\pi
T}\epsilon_{\mu\nu}V^*\partial_\nu V.
\end{equation}
The parameter $\mu$ is proportional to the fugacities of heavy
charged particles -- $W$ bosons and winos:
\begin{equation}
\mu\propto \exp\left(-{M_W\over T}\right).
\end{equation}

The 2D models of the type of Eq.~(\ref{comact}) have been
extensively studied in the literature starting with~\cite{jose}.
For a recent discussion see~\cite{dorey}. For $T<{g^2\over 8\pi}$,
the last term in the action, which contains the Polyakov loops, is
irrelevant, and can be neglected in discussing the infrared
physics. At these low temperatures the photon self-interaction
term $V^4+c.c.$ is relevant, and the vortex operator has a non
vanishing expectation value. At $T=T_c={g^2\over 8\pi}$ it becomes
irrelevant. If not for the Polyakov loop, the theory would be in a
massless phase above this temperature with the
Berezinsky-Kosterlitz-Thouless transition between the
phases~\cite{bkt}. Just like in the GG model, the transition into
the massless phase corresponds to logarithmic binding of monopoles
(or rather monopole pairs in the present theory) into
molecules~\cite{nk}. However, the Polyakov loop becomes relevant
precisely at the same temperature $T_c$ and renders the theory
massive also in the high-temperature phase. The phase transition
at $T_c$ remains a continuous one. The critical conformal theory
has the central charge $c=1$ and is the theory of one massless
scalar field.

Note that as opposed to the GG model where the value of the
critical temperature depends on the Higgs mass~\cite{2, sk}, in
the present case $T_c$ appears to be independent of the Higgs mass
with exponential accuracy. The difference is in the fact that in
the GG case both the monopole and charge induced operators where
relevant at the transition point. The value of the temperature was
determined by the equality of (renormalized) monopole and charge
fugacities \cite{sk}. In the SUSY case, however, both operators
are irrelevant at $T_c$, and the values of respective fugacities
are of no importance as long as they are small.

The transition at $T_c$ is clearly a deconfining transition. The
magnetic $Z_2$ symmetry is restored, and the expectation value of
the vortex operator vanishes, $\langle V\rangle=0$. It is
interesting that the parity is also restored at the same point.
Above the transition, the monopole-induced photon self-interaction
term is irrelevant. Thus, the infrared physics at $T>T_c$ is
described by the Lagrangian
\begin{equation} S=  \int d^2x\left[\frac{T}{2g^2}(\partial_i\tilde\chi)^2
- \mu\left(P^2+P^{*2}\right) \right].\label{hight}\end{equation}

This is exactly the same as in the nonsupersymmetric
model~\cite{gg2}. The order parameter for parity is $\langle
V^2\rangle+c.c.$. The average $\langle V^2\rangle$ was calculated
in the first paper of~\cite{gg2} and found to be nonzero. One
could therefore suspect that in the present case this average is
also non vanishing and the parity remains broken at high
temperature. This is however not the case. As explained
in~\cite{gg2}, to calculate $\langle V^2\rangle$ it is not
sufficient to consider the infrared limit~(\ref{hight}), but is
rather necessary to include also the monopole contributions. In
fact, this particular expectation value is dominated completely by
the contribution of the one-monopole sector. The reason is that
the insertion of $V^2$ into the path integral induces an
(anti)vortex of the field $P$. Due to the Debye mass term
$P^2+c.c.$, the vortex is accompanied by a string and has a
linearly infrared divergent action. In the GG model, a monopole
also creates a vortex of $P$ and thus the string emanating from
$V^2$ can end on it. The one-monopole sector thus has a finite
action in the presence of $V^2$ and dominates the expectation
value. The situation in the present case is quite different in
this respect. A single monopole gives a vanishing contribution to
the partition function due to the photino zero modes. The only
contributions to partition function one can consider are those
originating from the action~(\ref{comact}) in expansion in powers
of $\bar\zeta^2$. Those are contributions of even number of
monopoles and are obviously equivalent to insertions of integer
powers of $V^4$. Each such insertion creates two rather than one
strings, and thus cannot screen a single insertion of $V^2$. We
thus conclude that all contributions to $\langle V^2\rangle$ have
linearly infrared divergent action and thus in the thermodynamic
limit $\langle V^2\rangle=0$. The parity symmetry is therefore
restored at all temperatures above $T_c$.

To summarize, we find that in the present model confinement
disappears and parity is restored at the same critical temperature
$T_c=g^2/8\pi$. The phase transition is the same as in the
$Z_4$-invariant spin model. In fact, the dimensionally-reduced
theory~(\ref{comact}) has $Z_4$ global symmetry rather than the
$Z_2\otimes Z_2$ magnetic symmetry plus parity. The action
eq.(\ref{comact}) also has a separate parity symmetry under which
the vortex field does not transform. The reason for this symmetry
enhancement is that the only degrees of freedom which couple the
parity transformation with part of the $Z_4$ group ($V\rightarrow
iV$) are photinos. In the absence of fermions, the original
Lagrangian~(\ref{action}) indeed has the full $Z_4$ symmetry
supplemented by parity. At high temperature, where the reduced
theory~(\ref{comact}) is valid, the fermions are "heavy", or
better to say all their correlation functions are short range.
Thus, they indeed disappear from the infrared theory, and the
symmetry is effectively enhanced. It is due to this symmetry
enhancement that the deconfining and parity restoring transitions
happen at the same temperature. While this is an interesting
phenomenon, it seems somewhat non generic. In particular, in
(3+1)D gauge theory with adjoint fermions there is no reason to
expect the deconfining and chiral symmetry restoring phase
transitions to coincide. The physical order parameter for
deconfinement is the 't~Hooft loop $V$ \cite{thooftloop}, while
for chiral symmetry it is the fermionic bilinear form
$\bar\lambda\lambda$. In 3+1 dimensions, the two have very
different nature. While $\bar\lambda\lambda$ is a local field, $V$
is a string-like object. It is thus difficult to imagine the two
combining into a single order parameter as it is the case in the
(2+1)D theory discussed in the present note. The lattice results
indeed suggest that at least in the $SU(3)$-theory in 3+1
dimensions the two transitions are distinct~\cite{adjoint}.

\section*{Acknowledgments}
This work was initiated during the visit of D.A. to the
Particle Theory Group at the University of Plymouth, and he is grateful to this Group
for a warm hospitality extended to him during this visit. He is also grateful for the permanently cordial hospitality
to the Theory Group of the Physics Department of the Pisa University.
The work of D.A. was supported by INFN and partially by the
INTAS grant Open Call 2000, project No. 110.
The work of
A.K. is supported by PPARC advanced fellowship. A.K. thanks David
McMullan for a useful conversation.


\newpage


\begin{thebibliography}{100}

\bibitem{2}
G.~Dunne, I.I.~Kogan, A.~Kovner, B.~Tekin, JHEP {\bf 01} (2001) 032 [{\tt hep-th/0010201}].


\bibitem{gg2} I.~Kogan, A.~Kovner, B.~Tekin, JHEP {\bf 03} (2001) 021
[{\tt hep-th/0101171}]; JHEP {\bf 05} (2001) 062 [{\tt hep-th/0104047}];
I.~Kogan, A.~Kovner, M.~Schvellinger,  JHEP {\bf 07} (2001) 019 [{\tt hep-th/0103235}].



\bibitem{gg3} I.~Kogan, A.~Kovner, B.~Tekin,
Phys. Rev.{\bf D 63} (2001) 116007 [{\tt hep-ph/0101040}].

\bibitem{kk} I.~Kogan and A.~Kovner, {\it Monopoles, vortices and strings:
confinement and deconfinement in 2+1 dimensions at weak coupling.}
To be published in: ``At the frontier of
particle physics, vol. 4'' (Ed. M.~Shifman) [{\tt hep-th/0205026}].

\bibitem{dkn} G.V.~Dunne, A.~Kovner, Sh.M.~Nishigaki,
Phys. Lett. {\bf B 544} (2002) 215
[{\tt hep-th/0207049}].


\bibitem{antonov} N.O.~Agasian and D.~Antonov, JHEP {\bf 06} (2001) 058 [{\tt hep-th/0104029}];
Phys. Lett. {\bf B 530} (2002) 153 [{\tt hep-th/0109189}]; D.~Antonov,
Phys. Lett. {\bf B 535} (2002) 236 [{\tt hep-th/0204114}];
Mod. Phys. Lett. {\bf A 17} (2002) 851 [{\tt hep-th/0204151}].


\bibitem{ahw}
I.~Affleck, J.~Harvey, E.~Witten, Nucl. Phys. {\bf B 206} (1982) 413.

\bibitem{MZ}
M.~Moshe and J.~Zinn-Justin,
Nucl. Phys. {\bf B 648} (2003) 131
[{\tt hep-th/0209045}].


\bibitem{1}A.M.~Polyakov, Nucl. Phys. {\bf B 120} (1977) 429.

\bibitem{kovner} A.~Kovner, {\it Confinement, magnetic Z(N) symmetry and low-energy effective theory of gluodynamics},
in: ``At the frontier of
particle physics, vol. 3'' (Ed. M.~Shifman) [{\tt hep-ph/0009138}].

\bibitem{jose} J.V.~Jos\'{e}, L.P.~Kadanoff, S.~Kirkpatrick, D.R.~Nelson, Phys. Rev. {\bf B 16} (1977) 1217.

\bibitem{dorey}P.~Dorey, R.~Tateo, K.E.~Thompson,
Nucl. Phys. {\bf B 470} (1996) 317
[{\tt hep-th/9601123}].

\bibitem{bkt}
V.L.~Berezinsky, Sov. Phys.- JETP {\bf 32} (1971) 493; J.M.~Kosterlitz and D.J.~Thouless,
J. Phys. {\bf C 6} (1973) 1181; J.M.~Kosterlitz, J. Phys. {\bf C 7} (1974) 1046.

\bibitem{nk}
N.O.~Agasyan and K.~Zarembo, Phys. Rev. {\bf D 57} (1998) 2475.
[{\tt hep-th/9708030}].

\bibitem{sk} Y.V.~Kovchegov and D.T.~Son, JHEP {\bf 01} (2003) 050 [{\tt hep-th/0212230}].

\bibitem{thooftloop} C.~Korthals-Altes and A.~Kovner,  Phys. Rev. {\bf D 62} (2000) 096008
[{\tt hep-ph/0004052}].

\bibitem{adjoint} F.~Karsch and  M.~Lutgemeier, Nucl. Phys. {\bf B 550} (1999) 449
[{\tt hep-lat/9812023}].







\end{thebibliography}
\end{document}